\documentclass[letterpaper,twocolumn,10pt]{article}
\usepackage{usenix2019_v3}

\usepackage[table]{xcolor}

\usepackage{tikz}
\usepackage{amsmath}
\usepackage[normalem]{ulem}

\usepackage{filecontents}

\usepackage{xurl} 
\usepackage{multicol, multirow, tabularx}
\usepackage{url}
\usepackage{csquotes}
\usepackage[normalem]{ulem}
\usepackage{tcolorbox}
\usepackage{caption}
\usepackage[htt]{hyphenat}
\usepackage[hyphenbreaks]{breakurl}
\usepackage[export]{adjustbox}
\usepackage{amssymb}

\usepackage{csquotes}
\usepackage{threeparttable}
\usepackage{subcaption}
\usepackage{seqsplit}
\usepackage[english]{babel}
\usepackage{CJKutf8}

\begin{document}

\date{}


\title{\Large \bf RiskHarvester: A Risk-based Tool to Prioritize Secret Removal Efforts in Software Artifacts}





\author{\rm Setu Kumar Basak\qquad Tanmay Pardeshi\qquad Bradley Reaves\qquad Laurie Williams\\\rm North Carolina State University, USA}




\maketitle

\begin{abstract}
Since 2020, GitGuardian has been detecting checked-in hard-coded secrets in GitHub repositories. During 2020-2023, GitGuardian has observed an upward annual trend and a four-fold increase in hard-coded secrets, with 12.8 million exposed in 2023. However, removing all the secrets from software artifacts is not feasible due to time constraints and technical challenges. Additionally, the security risks of the secrets are not equal, protecting assets ranging from obsolete databases to sensitive medical data. Thus, secret removal should be prioritized by security risk reduction, which existing secret detection tools do not support. \textit{The goal of this research is to aid software practitioners in prioritizing secrets removal efforts through our security risk-based tool}. We present RiskHarvester, a risk-based tool to compute a security risk score based on the value of the asset and ease of attack on a database. We calculated the value of asset by identifying the sensitive data categories present in a database from the database keywords in the source code. We utilized data flow analysis, SQL, and Object Relational Mapper (ORM) parsing to identify the database keywords. To calculate the ease of attack, we utilized passive network analysis to retrieve the database host information. To evaluate RiskHarvester, we curated RiskBench, a benchmark of 1,791 database secret-asset pairs with sensitive data categories and host information manually retrieved from 188 GitHub repositories. RiskHarvester demonstrates precision of (95\%) and recall (90\%) in detecting database keywords for the value of asset and precision of (96\%) and recall (94\%) in detecting valid hosts for ease of attack. Finally, we conducted a survey (52 respondents) to understand whether developers prioritize secret removal based on security risk score. We found that 86\% of the developers prioritized the secrets for removal with descending security risk scores.
\end{abstract}

\section{Introduction} \label{Introduction}
In March 2024, GitGuardian reported an increasing trend in the past four years on secrets exposure in GitHub repositories~\cite{gitguardian-secret-sprawl}. In 2023 alone, 12.8 million new secrets were exposed, a four-fold increase compared with 2020. Of the 14.9 million developers who pushed code in GitHub, 1.7 million exposed secrets, such as API keys and database passwords, essential for connecting to external services~\cite{gitguardian-secret-sprawl}. Developers keep hard-coded secrets in application packages and version control systems~\cite{meli2019bad}, leaving sensitive data vulnerable, as seen in Uber's 2022 breach~\cite{uber-breach}, where attackers exploited secrets in a PowerShell script to access Uber's internal tools. 

However, removing all the secrets from software artifacts is not feasible. Rayhanur et al.~\cite{rahman2022secret} found developers ignoring secret detection tool alerts due to false positives, time pressure, and technical challenges. Additionally, the security risks of the secrets are not equal, protecting assets ranging from obsolete databases to sensitive medical data. Thus, secret removal should be prioritized by security risk reduction, which existing secret detection tools do not support. Developers may stop using the tools due to ``alert fatigue''~\cite{alert-fatigue} if the alerts are not efficiently prioritized for secret removal. 

Existing secret detection tools prioritize secret removal based on ``severity'', a rating tied to the secret type~\cite{ggshield-severity}. For example, the same severity rating is assigned to any database secret without considering the protected asset information. However, the value of the asset protected by the secret can vary from a database with mock data to medical data whose breach can incur fines. Similarly, the ease of accessing an asset varies. For example, an asset with a public IP address can be easy for attackers to access. In contrast, an asset with a private IP address or localhost will require attackers to be on the same network or have physical access to the host machine. 

The National Institute of Standards and Technology (NIST) defines the security risk of an entity as a function of the impact of the adverse event and the likelihood of the event by a threat source~\cite{nist-risk}. For a secret, the security risk can be defined as the function of the value of asset and the attacker's ease of accessing the asset protected by the secret. Protection Poker~\cite{williams2010protection}, a threat modeling game, employs the product of relative measures of ``value points'' and ``ease points'' for security risk quantification, such as one requirement being five times easier to attack than another. Similarly, for a secret, the security risk can be defined as the \textit{product of value of asset and ease of attack}. This security risk computation is based upon the hypothesis that attackers are more likely to succeed in attacking assets of high value and that are easier to attack. Table~\ref{protection-poker-relative-estimation} provides an example of security risk computation for three secret-asset pairs. A secret-asset pair consists of a secret, such as a database password, and a protected asset by the secret, such as the database server. Pair 3 is deemed to have the highest security risk because the value of the asset is 40 times more valuable than Pair 1 and 100 times easier to attack than Pair 2. Thus, removing the Pair 3 secret from the source code is of primary importance. We hypothesize that providing a security risk score for each secret can aid developers in prioritizing the secret removal efforts. \textit{The goal of our study is to aid software practitioners in prioritizing secrets removal efforts through our security risk-based tool.}

\begin{table}[!t]
\centering
\small
\caption{Security Risk for Each Secret-Asset Pair}
\label{protection-poker-relative-estimation}
\begin{tabular}{|ll|l|l|l|}
\hline
\multicolumn{2}{|c|}{\textbf{\begin{tabular}[c]{@{}c@{}}Secret-\\ Asset Pair\end{tabular}}} &
  \multicolumn{1}{c|}{\textbf{\begin{tabular}[c]{@{}c@{}}Value of \\ Asset\end{tabular}}} &
  \multicolumn{1}{c|}{\textbf{\begin{tabular}[c]{@{}c@{}}Ease of \\ Attack\end{tabular}}} &
  \multicolumn{1}{c|}{\textbf{Security Risk}} \\ \hline
\multicolumn{2}{|l|}{\textbf{Pair 1}} & 1  & 100 & 100   \\ \hline
\multicolumn{2}{|l|}{\textbf{Pair 2}} & 40 & 1   & 40   \\ \hline
\multicolumn{2}{|l|}{\textbf{Pair 3}} & 40 & 100 & 4000 \\ \hline
\end{tabular}%
\end{table}

In this research, we studied how we can programmatically calculate the security risk score by identifying the value of asset and ease of attack for each secret-asset pair and provided answers to our research questions:

\textbf{RQ1:} What performance can be achieved in automatically identifying the value of asset and ease of attack for secret-asset pairs in terms of precision, recall, and F1 score? (Section~\ref{RiskHarvesterResult})

\textbf{RQ2:} Do developers prioritize secret removal based on the descending security risk scores? (Section~\ref{DeveloperSurvey})

We constructed RiskHarvester, a risk-based tool to provide security risk scores of database secret-asset pairs based on the value of asset and ease of attack. We calculated the value of asset by identifying the categories of sensitive data, such as personal information present in the database, from the database keywords (database, table, and column names). We utilized data flow analysis, SQL, and Object Relational Mapper (ORM) parsing to detect the database keywords from the source code. To calculate the ease of attack, we used passive network analysis to retrieve the database host information.

To answer RQ1, we constructed RiskBench, a benchmark of 1,791 database secret-asset pairs from 188 GitHub repositories. We manually inspected each secret-asset pair and included the database keywords, corresponding sensitive data categories, and valid host information in RiskBench. We evaluated RiskHarvester against RiskBench in identifying the database keywords, sensitive data categories, and valid hosts for each secret-asset pair. To answer RQ2, we conducted an online developer survey to understand whether developers use the descending security risk scores to prioritize secret removal from software artifacts. We hypothesize that developers will prioritize secret removal ranked by descending risk scores. We provided a summary of our contributions as follows:

\begin{itemize}
    \item We automatically computed relative security risk scores of checked-in secrets to aid developers in prioritizing secret removal, which existing secret detection tools do not support. Additionally, we reported the developer study findings on the effectiveness of the security risk score in the alerts for secret removal prioritization.
    \item We made the implementation of RiskHarvester publicly available~\cite{riskartifacts}. We also provided RiskBench, a dataset of secret-asset pairs to aid researchers and developers, available via a data protection agreement.
\end{itemize}

Our paper is organized as follows: Section~\ref{Methodology} depicts our research methodology. We discuss the RiskHarvester construction and evaluation results against RiskBench in Sections~\ref{RiskHarvester} and~\ref{Results}, respectively, followed by the implications and limitations of our work. We discuss the related work in Section~\ref{RelatedWork} and conclude in Section~\ref{Conclusion}, followed by ethics considerations and open science policy compliance.

\section{Research Methodology} \label{Methodology}
In this section, we explain the process of RiskBench curation, identifying the value of asset and ease of attack patterns for calculating security risk score, and the developer survey.

\subsection{RiskBench Curation} \label{RiskBench}

To select a dataset of secret-asset pairs for calculating security risk score, we started with AssetBench~\cite{assetharvester}, a publicly-available dataset of secret-asset pairs. We accessed the dataset through Google Cloud (ID: \seqsplit{dev-range-332204.assetbench}). The authors of AssetBench curated 818 repositories from the September 2022 snapshot of GitHub's Google BigQuery Dataset (ID: bigquery-public-data.github\_repos)~\cite{google-big-query}. The dataset contains 97,479 manually labeled secrets (as true or false), extracted using two open-source secret detection tools, TruffleHog~\cite{trufflehog} and Gitleaks~\cite{gitleaks}. In addition, two authors of AssetBench manually inspected candidate asset-containing files to identify the assets protected by the corresponding secrets. The dataset also provides metadata such as repository name, commit ID, file path, and the line number where the secret-asset pairs have been identified. However, the dataset does not contain the database keywords (database, table, and column names) and corresponding sensitive data categories for each secret-asset pair. Additionally, the dataset lacks information on whether the asset identifier is a placeholder. In our study, we have utilized the database keywords, data categories, and asset identifiers for calculating the value of asset and ease of attack (Section~\ref{RiskHarvester}). Thus, to evaluate the performance of RiskHarvester in identifying the value of asset and ease of attack for each secret-asset pair (RQ1), we extended the dataset as \textit{RiskBench} by including the additional information.

\textbf{Filtering Dataset}: Before identifying the value of asset and ease of attack information for each secret-asset pair, we applied the following selection criteria to filter AssetBench. 

\textit{\uline{Criteria 1 (Programming Language):}} According to the 2023 GitGuardian report~\cite{gitguardian-secret-sprawl-2023}, developers most frequently exposed secrets in source code written in Python in GitHub. Thus, we selected repositories containing Python source code in our study. We selected 188 repositories from 818 repositories and 34,569 secrets from 97,479 secrets of AssetBench.

\textit{\uline{Criteria 2 (Secret Type):}} The 2024 GitGuardian report~\cite{gitguardian-secret-sprawl} reveals that out of 12.8 million exposed secrets in public GitHub repositories, the top secret type is database secrets. Thus, we selected database secret-asset pairs to calculate the security risk score in our study. However, we need to narrow the scope to maintain our study's feasibility since multiple database providers are present. We observed that according to the Stack Overflow Developer Survey 2024~\cite{SOdevelopersurvey2024}, the top five databases used by developers are PostgreSQL~\cite{postgresql}, MySQL~\cite{mysql}, SQLite~\cite{sqlite}, SQL Server~\cite{sqlserver}, and MongoDB~\cite{mongodb}. However, we excluded SQLite since SQLite is a file-based database requiring no authentication. Finally, we filtered the secret-asset pairs of the four databases and selected 1,791 secret-asset pairs from 34,569 secrets.

Table~\ref{asset-types} shows the number of secret-asset pairs of the four database types with the percentage of each type. We observed that only 25 secret-asset pairs (1.4\% of the total) are present for SQL Server. The lower percentage may be due to SQL Server's proprietary nature, limiting the adoption in open-source projects compared to open-source databases.

\begin{table}[!t]
\centering
\caption{Count of Secret-Asset pairs in RiskBench}
\label{asset-types}
\small
\begin{tabular}{|c| c| c|} 
 \hline
 \textbf{Database Type} & \textbf{\# Secret-Asset Pair} & \textbf{\% of Pair} \\ [0.5ex] 
 \hline
 MySQL & 777 & 43.4\%\\ 
 \hline
 PostgreSQL & 679 & 37.9\%\\
 \hline
 MongoDB & 310 & 17.3\%\\
 \hline
 SQL Server & 25 & 1.4\%\\ 
 \hline
\end{tabular}
\end{table}

\textbf{Identifying Database Keywords, Sensitive Data Categories, and Asset Identifier Information}: To identify the database keywords, the first and second authors of the paper manually inspected each secret-asset pair using the repository name, commit ID, file path, and line number provided by the dataset. Since the keywords, such as table and column names, may not be present in the same file where the asset is located, both authors inspected the candidate database keywords containing files in the repository. Finally, the database name, corresponding table, and column names are collected.

Next, to assign a sensitive data category for each database keyword, we utilized the data categories provided by Google Cloud Data Loss Prevention (DLP)~\cite{google-dlp}. The Google Cloud DLP is a service that helps organizations discover and classify sensitive data to comply with GDPR, HIPAA, and PCI-DSS regulations~\cite{google-dlp}. The Google Cloud DLP provides 192 sensitive data categories grouped into 7 domains. These 7 domains include Personally Identifiable Information (PII), such as a name, and Sensitive Personally Identifiable Information (SPII), such as a passport number. In addition, each data category is assigned a sensitivity level (``HIGH'', ``MODERATE'' and ``LOW''). Table~\ref{sensitive-data-types} presents an example of a data category and the corresponding sensitivity level in each domain. However, we observed that data categories contain similar information. For example, data categories such as ``CANADA\_PASSPORT'' and ``US\_PASSPORT'' contain passport information, and these data categories have the same sensitivity level. Since we will compute the similarity score of a database keyword with the data category (Step 2.2, Section~\ref{RiskHarvester}) to assign the correct data category, reducing the number of comparisons will improve the mapping performance. Thus, we manually inspected each data category and merged similar categories into a generic data category such as ``PASSPORT''. Finally, we identified 113 data categories. Both authors independently assigned a data category to each database keyword.   


\begin{table}[!t]
\centering
\caption{Examples of a data category with the corresponding sensitivity level for seven domains provided by Google Cloud DLP. The full list can be found online~\cite{riskartifacts}.}
\label{sensitive-data-types}
\small
\begin{tabular}{|l|l|l|}
\hline
\textbf{Domain} & \textbf{Data Category} & \textbf{Sensitivity} \\ \hline
PII                    & PERSON\_NAME       & MODERATE             \\ \hline
SPII                   & PASSPORT           & HIGH                 \\ \hline
DEMOGRAPHIC            & GENDER             & MODERATE             \\ \hline
CREDENTIAL             & AUTH\_TOKEN        & HIGH                 \\ \hline
GOVERNMENT\_ID         & VAT\_NUMBER        & HIGH                 \\ \hline
DOCUMENT               & RESUME             & MODERATE             \\ \hline
\begin{tabular}[c]{@{}l@{}}CONTEXTUAL\_\\ INFORMATION\end{tabular} & \begin{tabular}[c]{@{}l@{}}ORGANIZATION\_\\ NAME\end{tabular} & LOW \\ \hline
\end{tabular}%
\end{table}

Next, both authors inspected the asset identifier for a secret and labeled whether the database host is a placeholder considering the asset source code context. The agreement of finding the database keywords, corresponding data categories, and if the host is a placeholder with a Cohen's Kappa~\cite{cohen-kappa} score of 0.88, 0.93, and 0.85, respectively, between the two authors. These scores indicate a ``near perfect agreement'' according to Landis and Koch's interpretation~\cite{landis-koch}. The disagreements were resolved after a discussion between the two authors.

\subsection{Value of Asset and Ease of Attack Patterns} \label{Patterns}

For each secret-asset pair, we calculated the security risk score as the product of the value of asset and ease of attack, as defined in Equation~\ref{eqriskscore}. This security risk computation is based upon the hypothesis that attackers are more likely to succeed in attacking assets of high value and that are easier to attack.

\begin{gather} \label{eqriskscore}
    \begin{aligned}[b]
        \textnormal{Security Risk = (Value of Asset) x (Ease of Attack)}
   \end{aligned}  
\end{gather}

\begin{figure}[!t]
    \centering
    \begin{subfigure}[b]{\columnwidth}
        \centering
        \includegraphics[width=\columnwidth, frame]{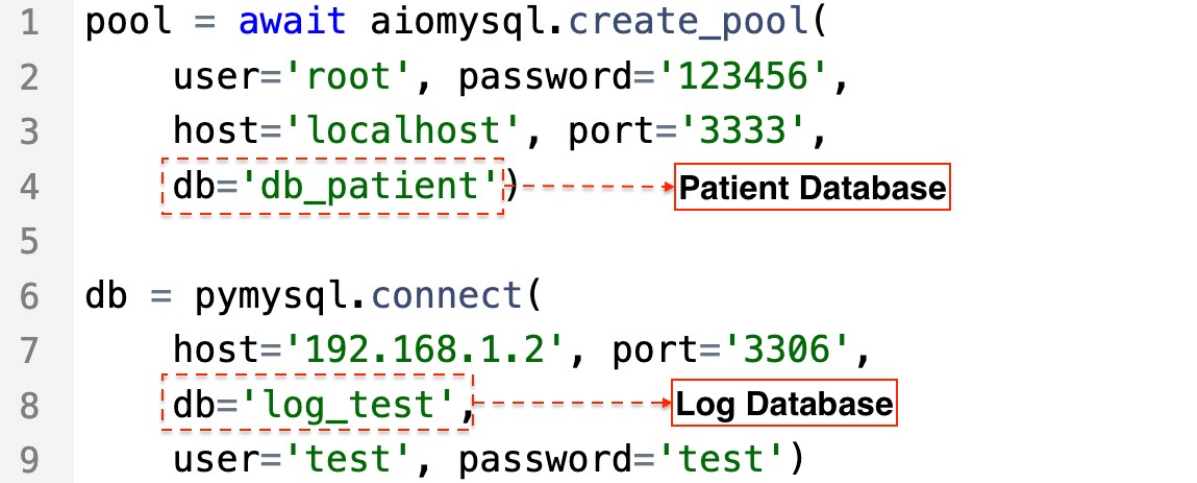}
        \caption[]{{The value of asset can be inferred from the database name}}
        \label{fig:only-db-name}
    \end{subfigure}
\vskip\baselineskip
    \begin{subfigure}[b]{\columnwidth}
        \centering
        \includegraphics[width=\columnwidth, frame]{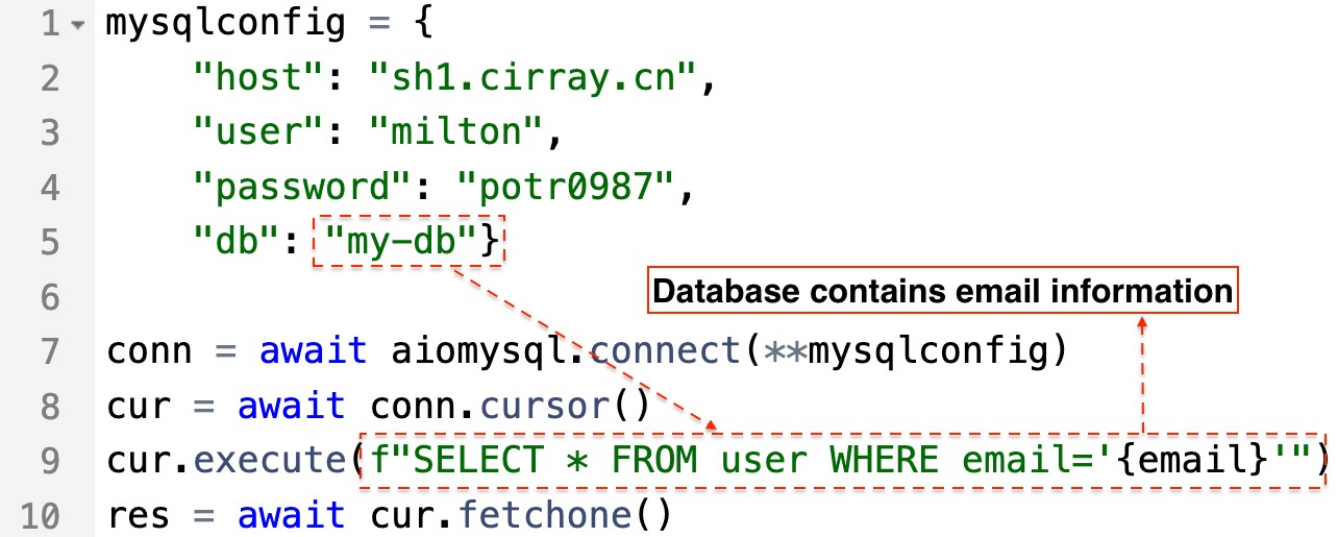}
        \caption[]{{The value of asset can be inferred from table and column names}}
        \label{fig:table-column-name}
    \end{subfigure}
\caption{Asset's value can be inferred from the database name, table names, and column names from the source code.}
\label{fig:value-of-asset-infer}
\end{figure}

\begin{figure*}[!ht]
    \centering
    \begin{minipage}{0.49\textwidth} 
        \centering
        \begin{subfigure}[b]{\textwidth}
            \centering
            \includegraphics[width=\textwidth, frame]{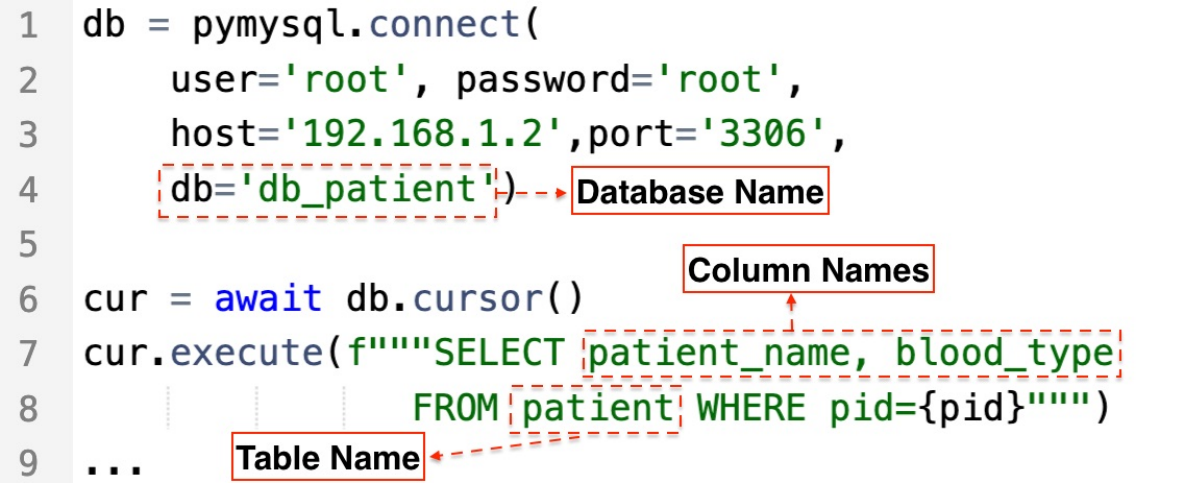}
            \caption[]{{V-Pattern 1 (SQL Database Driver Calls)}}
            \label{fig:vpattern1}
        \end{subfigure}
        \vskip\baselineskip 
        \begin{subfigure}[b]{\textwidth}
            \centering
            \includegraphics[width=\textwidth, frame]{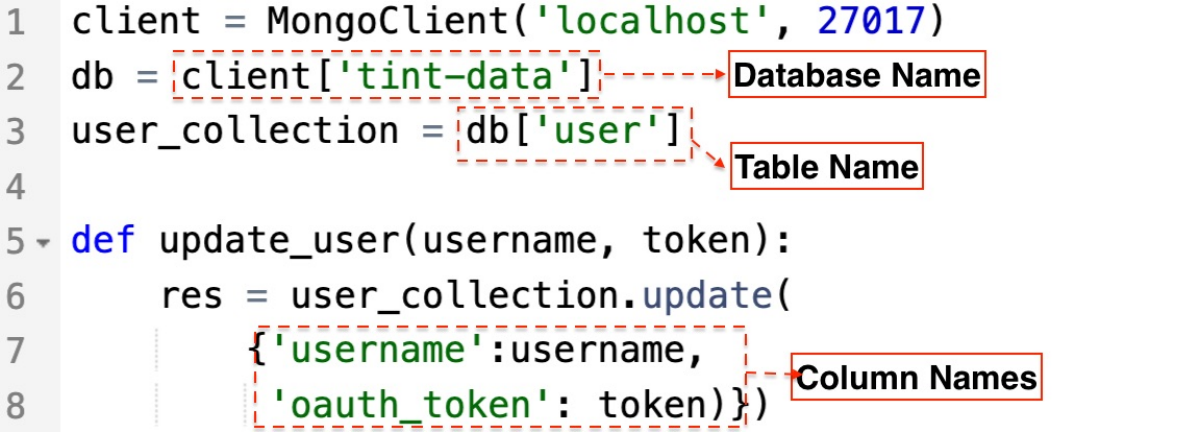}
            \caption[]{{V-Pattern 2 (NoSQL Database Driver Calls)}}
            \label{fig:vpattern2}
        \end{subfigure}
    \end{minipage}
    \hfill
    \begin{minipage}{0.49\textwidth} 
        \centering
        \begin{subfigure}[b]{\textwidth}
            \centering
            \includegraphics[width=\textwidth, frame]{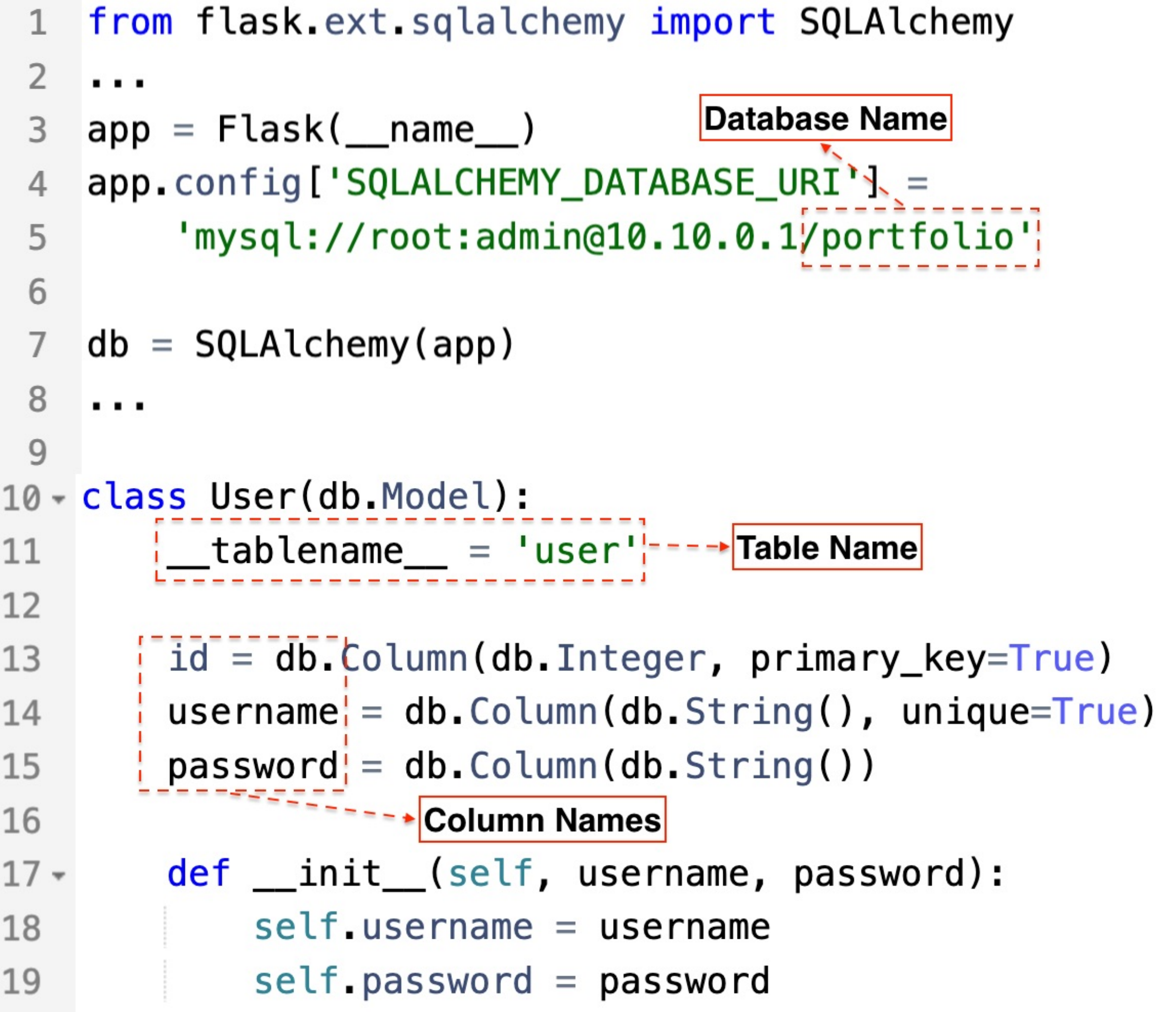}
            \caption[]{{V-Pattern 3 (ORM Framework Calls)}}
            \label{fig:vpattern3}
        \end{subfigure}
    \end{minipage}
\caption{We identified three patterns to locate database, table, and column names for each secret-asset pair in the source code.}
\label{fig:value-of-asset-patterns}
\end{figure*}

The first and second authors independently inspected a random sample of 50 secret-asset pairs from RiskBench and developed patterns for programmatically calculating the value of asset and ease of attack for each secret-asset pair. Now, we describe the observed patterns, which form the basis of RiskHarvester construction in calculating security risk score (Section~\ref{RiskHarvester}). The source code snippets of Figure~\ref{fig:value-of-asset-infer} and~\ref{fig:value-of-asset-patterns} are taken from RiskBench repositories of the 50 secret-asset pairs.

\textbf{Value of Asset Patterns}: A database asset identifier has three parts (host, port, and database name)~\cite{asset-identifier}. We observed that the asset's value can be inferred from the database name since the name reveals the type of data the database contains. Figure~\ref{fig:only-db-name} shows that a patient (\texttt{"db\_patient"}) and a test log (\texttt{"log\_test"}) database is passed in the \texttt{"db"} arguments where the value of patient database will be relatively higher than that of a test log database. However, we may not always be able to infer the database's data from the database name. Figure~\ref{fig:table-column-name} shows that the database name is \texttt{"my-db"} (line 5), posing difficulty in inferring the value of the asset. However, we observed that the database asset has been configured in line 7, and a SQL query is executed to access the user table containing the email column. Thus, we can retrieve the type of data of the database from the table and column names which we used to calculate the value of asset (Step 2.2, Section~\ref{RiskHarvester}). 

We now describe the three mutually exclusive patterns observed in the database, table, and column name locations for an asset identifier. The numbers in parentheses denote occurrences of each pattern in 50 secret-asset pairs.

\textit{\uline{V-Pattern 1 (SQL Database Driver Calls) (25)}}: We observed that the database name and the corresponding table and column names of relational databases can be found in the SQL database drivers, such as MySQL drivers~\cite{pymysql}. These SQL database drivers provide functions to connect to databases, execute queries, manage transactions, and fetch results. We observed that the database name is passed in the same function where the database secret and server address information is also passed to set up the connection. For example, Figure~\ref{fig:vpattern1} shows that the database name (\texttt{"db\_patient"}) is passed in \texttt{pymysql.connect} function (line 4). Additionally, the corresponding database table and column names are present in raw SQL queries passed in query functions of SQL database drivers such as the ``execute'' function (line 7, Figure~\ref{fig:vpattern1}). 

\textit{\uline{V-Pattern 2 (NoSQL Database Driver Calls) (16)}}: The database, table, and column names of non-relational databases can be found in the NoSQL database drivers, such as MongoDB drivers~\cite{mongodb-driver}. Unlike relational databases, non-relational databases are document or key-value pair databases without a structured schema. The table and column names are referred to as collection and field names, respectively. Figure~\ref{fig:vpattern2} shows that the database name and collection name are passed as a dictionary key to the NoSQL driver client and db instance in lines 2 and 3, respectively. However, the field names are passed as key-value pairs in a dictionary object instead of as a raw SQL string to the driver query function (lines 7-8), where each key is the field name of the corresponding collection.

\textit{\uline{V-Pattern 3 (ORM Framework Calls) (9)}}: We observed that the database name and the corresponding table and column names of relation databases can be found in Object Relational Mapper (ORM) framework calls~\cite{ORM}, such as SQLAlchemy~\cite{sqlalchemy}. Unlike other drivers, ORM abstracts database access through objects rather than directly managing the access with raw SQL queries. Figure~\ref{fig:vpattern3} shows that the database name (\texttt{"portfolio"}) is defined in a connection string along with secret and server address and passed to ORM configuration (line 4). Since ORM maps tables in a relational database to classes and rows to instances of those classes, the table name and column names can be found in these classes. The \texttt{"\_\_tablename\_\_"} attribute defines the table name (line 11), and the other attributes define the column names, such as username and password (lines 14-15), as shown in Figure~\ref{fig:vpattern3}.

\textbf{Ease of Attack Patterns}: Similar to the value of an asset, ease of accessing the asset can vary based on multiple factors. For example, attackers can more easily access an asset with a public IP address, whereas attackers can not directly access an asset on a localhost. Since the server address is defined as a combination of host and port, we can infer the ease of accessing the asset from these parts of the asset identifier. 

Now, we describe the observed four patterns in the asset identifiers in the source code. The numbers in parentheses denote occurrences of each pattern in 50 secret-asset pairs.

\textit{\uline{E-Pattern 1 (DNS Name) (27)}}: We observed that developers put a DNS name, such as (\texttt{"sh1.cirray.cn"}), in the host part of the asset identifier as a database server address in the source code. However, not all the DNS names are resolved to IP addresses due to non-existent domains, misconfigured DNS records, or expired domains. In addition, we observed invalid DNS names that are not valid according to DNS name format or a placeholder/dummy such as \texttt{"your-project-name.com"}. Thus, the ease of attack of a secret-asset pair differs if DNS is resolvable (Steps 3.1.1 and 3.1.2, Section~\ref{RiskHarvester}).

\textit{\uline{E-Pattern 2 (IP Address) (23)}}: Developers put IP addresses such as (\texttt{\seqsplit{"185.60.21.35"}}) instead of DNS name as database server addresses in the host part of the asset identifier. However, not all the IP addresses are routable addresses. For example, localhost address (\texttt{"127.0.0.1"}) or private IP addresses (\texttt{"192.168.1.1"}) pose more difficulty to external attackers to access the asset than public IP addresses. We also observed invalid or placeholder IP addresses such as \texttt{"x.x.x.x"} or \texttt{"0.0.0.0"} that attackers can not leverage to access the asset. Thus, the ease of attack of a secret-asset pair differs if the IP address is routable (Steps 3.1.3 and 3.1.4, Section~\ref{RiskHarvester}).

\textit{\uline{E-Pattern 3 (Scannable) (7)}}: Not all the public IP addresses that are either present directly in the host or resolved from the DNS names are scannable since IP addresses can be present behind firewalls or other security measures. Scannable means the discovery of active services through network scans. The IP addresses that are scannable are easier to attack than those of non-scannable IP addresses (Step 3.1.5, Section~\ref{RiskHarvester}).

\textit{\uline{E-Pattern 4 (Port Open) (4)}}: Developers include the database server's port number in the asset identifier, such as port 3306 for MySQL database. If the IP address is publicly accessible and the database port is open, the attackers can more easily access the database. In contrast, access is relatively difficult if the database port is closed or restricted (Step 3.1.6, Section~\ref{RiskHarvester}). We observed four ports open in the 50 secret-asset pairs that we inspected using Censys~\cite{censys}, an online service that provides the port information of a host.

\begin{table*}[!ht]
\setlength{\tabcolsep}{1.5pt}
\centering
\caption{The Developer Survey Questions. The full questionnaire of the online survey can be found online~\cite{riskartifacts}.}
\label{survey-ques}
\footnotesize
\begin{tabular}{|lll|l|}
\hline
  \multicolumn{3}{|c}{\textbf{Secret Alert Information}} & \multicolumn{1}{|c|}{\multirow{2}{*}{\textbf{Question Description}}} \\ \cline{1-3} 
  \multicolumn{1}{|c}{\textbf{Secret A}} &
  \multicolumn{1}{c}{\textbf{Secret B}} &
  \multicolumn{1}{c}{\textbf{Secret C}} &
  \multicolumn{1}{|c|}{}\\ \hline

  \multicolumn{1}{|l|}{\begin{tabular}[c]{@{}l@{}}Secret: "Fm)4dj"\\ Severity: CRITICAL\\Other Info: Repo and File Location\end{tabular}} &
  \multicolumn{1}{l|}{\begin{tabular}[c]{@{}l@{}}Secret: "123456"\\ Severity: CRITICAL\\Other Info: Repo and File Location \end{tabular}} &
  \begin{tabular}[c]{@{}l@{}}Secret: "123456"\\ Severity: CRITICAL\\Other Info: Repo and File Location\end{tabular} &
  \begin{tabularx}{0.57\columnwidth}{X}\textbf{Q1}: Based on the alert information, in what order would you prioritize the removal of the secrets? Did the "Severity" info help?\end{tabularx} \\ \hline
  \multicolumn{1}{|l|}{\begin{tabular}[c]{@{}l@{}}Asset: "127.0.0.1"\\Other Info: Asset Location\end{tabular}} &
  \multicolumn{1}{|l|}{\begin{tabular}[c]{@{}l@{}}Asset: "111.230.140.27"\\Other Info: Asset Location\end{tabular}} &
  \multicolumn{1}{|l|}{\begin{tabular}[c]{@{}l@{}}Asset: "120.77.222.217"\\Other Info: Asset Location\end{tabular}} &
  \begin{tabularx}{0.57\columnwidth}{X}\textbf{Q2}: Given the additional asset information, would you change your prioritization order to remove the secrets? Why?\end{tabularx} \\ \hline
  \multicolumn{1}{|l|}{\begin{tabular}[c]{@{}l@{}}Security Risk Score: 100\\ Value of Asset: HIGH \\ (Blockchain data)\\ Ease of Attack: VERY\_DIFFICULT \\ (Localhost)\end{tabular}} &
  \multicolumn{1}{l|}{\begin{tabularx}{0.23\textwidth}{X}Security Risk Score: 40\\ Value of Asset: LOW (Video URL and TimeStamp data)\\ Ease of Attack: DIFFICULT (Public IP but not reachable)\end{tabularx}} &
  \begin{tabularx}{0.23\textwidth}{X}Security Risk Score: 320\\ Value of Asset: MODERATE \\ (Phone and Email Address data)\\ Ease of Attack: DIFFICULT \\ (Public IP, reachable but database port 3306 is not open)\end{tabularx} &
  \begin{tabularx}{0.57\columnwidth}{X}\textbf{Q3}: Given the additional security risk score (value of asset and ease of attack) information, would you change your prioritization order to remove the secrets? Why?\end{tabularx} \\ \hline \hline
  \multicolumn{3}{|l|}{\begin{tabularx}{0.7\textwidth}{X}The value of the asset is calculated based on the database table and column names from the source code. The ease of attack is calculated based on the accessibility of the database server. Next, we calculated the security risk score by multiplying the value of an asset and the ease of attack.\end{tabularx}} &
  \begin{tabularx}{0.57\columnwidth}{X}\textbf{Q4}: (Optional) Do you agree with our means of calculating the security risk score? What suggestions do you have for improving the security risk score?\end{tabularx} \\ \hline
\end{tabular}%
\end{table*}

\subsection{Developer Survey} To answer RQ2, we conducted a survey to understand whether developers prioritize the removal of secrets consistent with our hypothesis (based on the descending security risk scores).

\textbf{Participant Selection}: To find survey participants, we selected the developers who committed the secret-asset pairs in RiskBench. We selected these developers since they have experience with software secrets. We observed that the same developer committed multiple secret-asset pairs in a repository. Thus, we identified unique 1,478 committers from the 1,791 secret-asset pairs of RiskBench. In addition, we filtered out committers having a noreply (\seqsplit{xxx@noreply.github.com}) or GitHub Actions bot email address~\cite{github-no-reply} and selected 1,282 committers. Finally, we randomly selected 500 committers to avoid selection bias~\cite{selection-bias} to participate in the online survey.

\textbf{Survey Design}: Table~\ref{survey-ques} presents the four questions of the developer survey. We provided three alerts of database secrets of GitHub repositories detected by TruffleHog~\cite{trufflehog}. For each alert, we provided the secret, the ``Severity'' level, and the repository and file location provided by TruffleHog. Based on the alert information, we asked developers in which order they would prioritize secret removal and why. We hypothesize that developers will choose Secret A first without considering the asset information since Secret A looks like an actual password. Next, we provided the asset identifier (IP address) protected by the database secret with file location in each alert. Then, we asked if developers would change the priority of secrets removal based on the asset identifier and why. We hypothesize that developers will change their priority and choose B and C since these secrets point to public IP addresses. Next, we provided the security risk score, value of asset, and ease of attack information, such as the sensitive data categories and passive network information from RiskHarvester (Section~\ref{RiskHarvester}). Then, we asked if developers would change the priority of secret removal based on the security risk score and why. We hypothesize that developers will change their order to Secret C, A, and B. We also asked an optional question on developers' suggestions for improving the security risk score calculation. In the survey, all the questions are kept open-ended to avoid bias from predefined options and explore diverse perspectives.

\textbf{Conducting Survey}: For conducting the survey, we leveraged the Qualtrics~\cite{qualtrics}, a popular online survey host. However, before conducting the main survey, we conducted a pilot survey with five security researchers from the anonymized lab. In the pilot survey, we provided an additional question for suggestions on survey improvement, including any unclear, irrelevant, or overly detailed aspects. We conducted the main survey in November and December 2024. We offered a \$20 Amazon gift card to five randomly selected participants if they wished to participate in the lottery. We discussed the IRB approval and other ethical considerations in Section~\ref{Ethics}.

\section{RiskHarvester Construction} \label{RiskHarvester}

We calculated the security risk score as the product of value of asset and ease of attack for each secret-asset pair (Equation~\ref{eqriskscore}). We utilized the identified value of asset and ease of attack patterns (Section~\ref{Patterns}) and constructed RiskHarvester to calculate the security risk score. We now discuss the four-step process of constructing RiskHarvester.

\subsection{Step 1: Identifying Secret-Asset Pairs}

Before we identify the value of asset and ease of attack of secret-asset pairs, we used the implementation source code of AssetHarvester~\cite{assetharvester}, an open-source static analysis tool, to detect secret-asset pairs in a repository (Steps 1.1-1.3). AssetHarvester demonstrates precision of (97\%), recall (90\%), and F1-score (94\%) in detecting secret-asset pairs in RiskBench. We extended AssetHarvester as RiskHarvester to calculate the security risk score for each secret-asset pair (Steps 2-4).

\textbf{Step 1.1 Pattern Matching}: In the source code, a secret and the corresponding asset can be present in database connection strings that follow a specific format for different database types. For example, MySQL, PostgreSQL, and MongoDB follow the same connection string format ([scheme://][user:password@]host:port/db). Thus, regular expressions (regex) are formulated to identify the connection strings by manually analyzing the database documentation. In addition, the capturing group~\cite{capturinggroup} feature of the regex is utilized to isolate the secret and the corresponding asset (host, port, and db name) from the connection string. Table~\ref{regex-database-appendix} in the Appendix presents the regexes, which are grouped into three groups based on the connection string format similarity. 

\textbf{Step 1.2 Data Flow Analysis}: A secret and the corresponding asset can be defined separately in variables and passed to database driver functions instead of defined in a connection string. For example, Figure~\ref{fig:vpattern1} shows that the database secret and the corresponding asset are passed to the driver functions (lines 1-3). The secret-asset pair is passed to the driver function as positional or keyword arguments~\cite{positionalarg}. A positional argument is passed based on the position in the argument list without specifying the parameter name, whereas a keyword argument is passed by explicitly specifying the parameter name, such as ``password'' or ``host'', without fixed order in the function. Table~\ref{database-driver-types} presents the list of Python database drivers and ORM frameworks with the supported argument type. Since the argument positions and names for a secret-asset pair are known in the driver functions, data flow analysis~\cite{khedker2017data} is leveraged to identify the secret-asset pair by analyzing the data flow graph (DFG). DFG is a directed graph where the secret-asset pair is the source that flows into the driver function arguments, which act as sinks. CodeQL~\cite{codeql}, an open-source source code analysis framework that provides the data flow graph computed from the repository source code, is used for data flow analysis to identify the secret-asset pair.

\begin{table}[!t]
\centering
\caption{List of Python database drivers and ORM frameworks with their supported arguments for secret-asset pairs}
\label{database-driver-types}
\small
\begin{tabular}{|cc|l|l|l|}
\hline
\multicolumn{2}{|c|}{\textbf{Category}} &
  \textbf{Driver Name} &
  \textbf{\begin{tabular}[c]{@{}l@{}}Pos.\\ Arg.\end{tabular}} &
  \textbf{\begin{tabular}[c]{@{}l@{}}Key.\\ Arg.\end{tabular}} \\ \hline
\multicolumn{1}{|c|}{\multirow{8}{*}{\textbf{\begin{tabular}[c]{@{}c@{}}SQL \\ Driver\end{tabular}}}} &
  \multirow{2}{*}{MySQL} &
  aiomysql~\cite{aiomysql} &
   &
  \checkmark \\ \cline{3-5} 
\multicolumn{1}{|c|}{} &                             & PyMySQL~\cite{pymysql}         & \checkmark & \checkmark \\ \cline{2-5} 
\multicolumn{1}{|c|}{} & \multirow{3}{*}{PostgreSQL} & aiopg~\cite{aiopg}           & \checkmark & \checkmark \\ \cline{3-5} 
\multicolumn{1}{|c|}{} &                             & asyncpg~\cite{asyncpg}         & \checkmark & \checkmark \\ \cline{3-5} 
\multicolumn{1}{|c|}{} &                             & psycopg2~\cite{psycopg2}        & \checkmark & \checkmark \\ \cline{2-5} 
\multicolumn{1}{|c|}{} & SQL Server                  & pymssql~\cite{pymssql}         &   & \checkmark \\ \cline{2-5} 
\multicolumn{1}{|c|}{} & ODBC                        & pyodbc~\cite{pyodbc}          & \checkmark &   \\ \cline{2-5} 
\multicolumn{1}{|c|}{} & JDBC                        & JayDeBeApi~\cite{jaydeapi}      & \checkmark &   \\ \hline
\multicolumn{1}{|l|}{\multirow{2}{*}{\textbf{\begin{tabular}[c]{@{}l@{}}NoSQL\\ Driver\end{tabular}}}} &
  \multicolumn{1}{l|}{\multirow{2}{*}{MongoDB}} &
  pymongo~\cite{pymongo} &
   &
  \checkmark \\ \cline{3-5} 
\multicolumn{1}{|l|}{} & \multicolumn{1}{l|}{}       & Flask-PyMongo~\cite{flask-pymongo}   &   & \checkmark \\ \hline
\multicolumn{2}{|c|}{\multirow{3}{*}{\textbf{\begin{tabular}[c]{@{}c@{}}ORM\\ Framework\end{tabular}}}} &
  peewee~\cite{peewee} &
  \checkmark &
  \checkmark \\ \cline{3-5} 
\multicolumn{2}{|c|}{}                               & SQLAlchemy~\cite{sqlalchemy}      & \checkmark & \checkmark \\ \cline{3-5} 
\multicolumn{2}{|c|}{}                               & Django~\cite{django}          &   & \checkmark \\ \hline
\end{tabular}%
\end{table}

\textbf{Step 1.3 Fast-Approximation Heuristics}: The data flow analysis may not always be captured when source code has dynamic behavior, such as extensive use of reflection. In such cases, the secret-asset pair can be identified from the neighboring lines in the source code. Secrets are first extracted using two open-source secret detection tools, TruffleHog~\cite{trufflehog} and Gitleaks~\cite{gitleaks}. Next, an IP address or DNS name is searched in the three neighboring lines of the secret. Since multiple assets can be present in the neighboring lines, the prefixes of both the secret and asset variables are matched to find the correct asset. For example, ``mysql'' is the prefix of MySQL database secret (``mysql-password'') and server (``mysql-host'') variables.


\subsection{Step 2: Identifying Value of Asset}
In this section, we described the process of extracting database keywords from the source code (Step 2.1) and mapped these identified keywords to sensitive data categories (Step 2.2).

\textbf{Step 2.1 Extracting Database Keywords}: We extracted the database keywords (database, table, and column names) from database drivers and ORM frameworks. Our study included eight SQL and two NoSQL database drivers and three ORM frameworks for extracting database keywords (Table~\ref{database-driver-types}).

\uline{SQL Database Driver Calls}: We observed that the database name and corresponding table and column names are passed to SQL database driver functions (V-Pattern 1). The database name is passed as a positional or keyword argument based on SQL driver types along with the secret, host, and port in the same driver function, such as in the \texttt{pymysql.connect} function (lines 1-4, Figure~\ref{fig:vpattern1}). Thus, we included the database name argument in the data flow analysis of Step 1.2 and identified the database name along with the host and port.

Additionally, we observed that raw SQL queries are passed in query functions other than the ``connect'' function where the secret-asset pair is passed. Figure~\ref{fig:vpattern1} shows a SELECT SQL query is directly passed in the ``execute'' function (lines 7-8) for retrieving the patient information. However, SQL queries can also be defined in separate files such as .sql and .ddl files, which are mostly used for database migration and executed from the source code. However, CodeQL does not support data flow between source codes of multiple file types. Thus, the flow of raw SQL present in the .sql file can not be captured into the Python database driver's ``execute'' function. As a result, we first identified the SQL file name from the file open functions~\cite{python-open} using data flow analysis. Finally, to parse the table and column names from the raw SQL, we used the \texttt{sql\_metadata}~\cite{sql-metadata} package of Python that provides a parser for retrieving table and column names from raw SQLs.


\uline{NoSQL Database Driver Calls}: We observed that the database name, corresponding table, and column names are passed in the NoSQL database drivers (V-Pattern 2) for non-relational databases. However, unlike SQL database drivers, database and table names are passed as dictionary keys to the connection client and corresponding database instance. Thus, we first located the data flow node in the DFG for the connection client instance (sink) initialized with the secret-asset pair and traced the source that flows into the specific sink to extract the database name. Using the identified database name, we located the data flow node of the corresponding database instance (sink) and repeated the process to identify the table name that flows in the database instance sink. Since the column names are passed as key-value pairs in a dictionary in the driver query function, we first traced the data flow node of the dictionary that flows into the table instance sink. Next, to find the column names, we extracted the keys from the key-value pairs of the dictionary. Finally, the database, table, and column names of non-relational databases are extracted.

\uline{ORM Framework Calls}: From V-Pattern 3, we observed that developers employ ORM frameworks to access relational databases. For ORM framework calls, we found that the database name is passed to the ORM configuration functions as a part of the connection string. Thus, we located the configuration function sink in the DFG and extracted the database name by tracing the flow of the connection string into the sink (similar to Step 1.2). However, ORM abstracts database access through objects instead of raw SQL queries or key-value pairs. The database tables are mapped to model classes, and the columns are mapped to the attributes of the classes. Thus, we first located the ORM class that uses the ORM database instance in the DFG. Then, we identified the class source code from the abstract syntax tree and extracted the attribute names of the class. To extract the attribute names, we used Python's \texttt{py\_models\_parser}~\cite{py-models-parser} package, which can parse the model classes and table definitions. Finally, we separated the table and column names from the corresponding attribute names of the ORM class as database keywords.

\textbf{Step 2.2 Mapping Database Keywords to Sensitive Data Categories}: Since each database keyword can have different sensitivity, we mapped each identified keyword to a data category of 113 categories provided by Google Cloud DLP (Section~\ref{RiskBench}). We observed that the Google Cloud DLP provides API to assign a data category to specific instances of the data. For example, instead of the database keyword ``passport'', the API takes a country-specific passport number as input and outputs the mapped data category. However, in our study, we only have the identified database keywords from the source code for secret-asset pairs (Step 2.1). In addition, we observed that the database keywords will not always match the data category name exactly. For example, the database keyword is ``NID\_NUMBER'', which should be assigned to the ``NATIONAL\_ID\_NUMBER'' category. We now discuss the lexical and semantic string similarity algorithms we used to map each database keyword to a data category.

\uline{Prefix Match}: We observed that database keywords match from the start of a data category. For example, the database keyword is ``FINANCIAL\_ACC'', and the corresponding data category is ``FINANCIAL\_ACCOUNT\_NUMBER''. To measure the similarity between these strings, we utilized the Jaro-Winkler algorithm~\cite{winkler1990string} that emphasizes prefix similarity by assigning higher scores to strings that share common prefixes. The algorithm generates a similarity score between 0 and 1, where 0 indicates entirely dissimilar strings, and 1 indicates identical strings. We set a cut-off score of 0.7. To employ the Jaro-Winkler algorithm, we leveraged the \texttt{jaro\_winkler\_similarity} function of Python's \texttt{jellyfish}~\cite{jellyfish-python} package.

\uline{Substring Match}: We observed that database keywords may not have a longer common prefix with a data category. For example, the database keyword ``NID\_NUMBER'' should match the ``NATIONAL\_ID\_NUMBER'' category, though the middle characters are missing in the keyword. To address the scenario, we used the Gestalt pattern matching algorithm~\cite{black2004ratcliff}, which calculates a similarity by identifying the common substring and recursively comparing characters in the unmatched regions on both sides of the longest common string. Thus, we could match the database keyword with the correct category even if the database keyword is incomplete or has missing segments. Like the Jaro-Winkler algorithm, Gestalt provides a similarity score between 0 and 1, and we set a cut-off score of 0.7. We implemented the algorithm using the \texttt{SequenceMatcher} function of Python's \texttt{difflib}~\cite{sequencematcher} package.

\uline{Semantic Match}: We observed that database keywords differ lexically from the correct data category but have the same meaning. For example, the database keyword ``CELL\_NO'' should map to the ``PHONE\_NO'' category due to the same meaning. Thus, we need to calculate semantic similarity between the strings instead of lexical similarity (Prefix and Substring match). For semantic similarity between words, we leveraged fastText~\cite{bojanowski2017enriching}, a natural language processing (NLP) model for generating word embeddings by capturing semantic information. In addition, we observed that the subwords in the database keyword can be the same as the subwords of the data category but present in different orders. For example, despite the subword's order, the database keyword ``DATE\_OF\_BIRTH'' should match the ``BIRTH\_DATE'' category. We chose fastText over other NLP models, such as Word2Vec~\cite{mikolov2013efficient} and GloVe~\cite{pennington2014glove}, since fastText supports out-of-vocabulary word embeddings and is trained with character n-grams. As a result, fastText can be used to capture the similarity of the words with different subword orders. In our study, we used the pre-trained fastText model \texttt{cc.en.300.bin}, trained on Common Crawl and Wikipedia with 5-character n-grams, a window size of 5, and 10 negatives. We used the \texttt{fasttext}~\cite{fasttext} package of Python to access the model and calculate semantic similarity. We set a cut-off similarity of 0.65.

\uline{Non-English \& Transliterated Word Match}: We observed that non-English or transliterated words are present in the source code as database keywords. A transliterated word is a word from one language written in another language's alphabet by representing the pronunciation. For example, the Chinese word ``\begin{CJK}{UTF8}{gbsn}性别\end{CJK}'' or the corresponding transliterated word ``Xìngbié'' is present in the SQL queries. As a result, we first translated the non-English and transliterated words to English words and then computed the lexical and semantic similarity. In our study, we leveraged the Google Cloud's Translation API~\cite{google-cloud-translation-api} using the \texttt{google-cloud-translate} package~\cite{google-cloud-translate}.    

The cut-off similarity scores are chosen by randomly sampling database keywords and observing the score. We assigned a sensitivity level of ``UNSPECIFIED'' when no data category is matched, such as the database keyword ``test''.

\subsection{Step 3: Identifying Ease of Attack}

In this section, we described the process of identifying ease of attack information (Step 3.1) and assigning ease of attack categories based on the identified information (Step 3.2).

\textbf{Step 3.1 Finding Ease of Attack Information}: We identify different ease of attack information from the host and port part of the asset identifier after each step (Steps 3.1.1-3.1.6).

\uline{Step 3.1.1 Valid DNS Name}: From E-Pattern 1, we observed that developers put a DNS name in the host part of the asset identifier as a database server address. However, the DNS name can be invalid according to the DNS name format set by the Internet Engineering Task Force (IETF)~\cite{ietf} through Request for Comments (RFCs)~\cite{rfc}. For example, each segment between dots in the DNS name can have up to 63 characters and should not start or end with a hyphen. In our study, we utilized the \texttt{domain} function of Python's \texttt{validators}~\cite{validators} package to validate the DNS name format. 

Additionally, developers can put a placeholder DNS name in the source code, such as \texttt{"www.example.com"}. However, detecting placeholder DNS names is challenging because the placeholder DNS names can conform to the DNS name format, and no universal registry exists to identify them. We can apply a rule-based approach by analyzing the common placeholder keywords to detect placeholder DNS names. However, the rule-based approach has limitations, such as a lack of adaptability due to a fixed set of keywords to arbitrary DNS names. Additionally, the rule-based approach cannot interpret the meaning behind names. However, we can apply Large Language Models (LLMs) to detect the placeholder DNS names since LLMs excel in semantic understanding and recognizing contextual clues that differentiate actual DNS names from placeholders~\cite{liu2023summary, yang2024harnessing}. While other Generative pre-trained transformer (GPT) style LLMs exist, we leveraged ChatGPT due to ChatGPT's performance in Zero-shot Learning (ZSL) through Chain-of-Thought (CoT) prompting~\cite{yang2024harnessing, wei2022chain, brown2020language}. The ZSL enables models to address unseen tasks without prior training examples, while CoT prompting guides the models through a structured, step-by-step reasoning process to arrive at more accurate answers. In our study, we employed \texttt{gpt-4o-2024-08-06}~\cite{gpt-model} model of ChatGPT with temperature 0.2 to make the model more deterministic and confident. As shown in Table~\ref{dns-prompt} of the Appendix, we provided one example of a placeholder and one example of actual DNS names with the context source code in the CoT system prompt. In the user prompt, we provided the DNS name to be classified as a placeholder or not, along with two neighboring lines of source code for context. Finally, we identified the valid DNS names from the prompt answer, which we passed on to the next step.

\uline{Step 3.1.2 Resolvable DNS Name}: We observed that all valid DNS names may not resolve to IP addresses due to non-existent domains or misconfigured DNS records (E-Pattern 1). Thus, we checked whether the DNS names from Step 3.1.1 are resolvable by querying the DNS servers. We leveraged \texttt{nslookup}~\cite{nslookup}, a BIND name server software member that obtains the mapping between a domain name and IP address. However, we observed that nslookup can return a Canonical Name (CNAME) record when queried for a DNS name. The DNS system allows aliases using CNAME records to simplify domain management, enabling a single canonical domain to represent multiple aliases. Thus, to identify the IP address from the A (IPv4) or AAAA (IPv6) record for the DNS name, we recursively queried using the canonical domain name. In our study, we used Python's \texttt{nslookup}~\cite{python-nslookup} package.

\uline{Step 3.1.3 Valid IP Address}: From E-Pattern 2, we observed that invalid or placeholder IP addresses are present in the source code. To check the validity of the IP address directly present in the host part or the resolved IP address from the DNS name (Step 3.1.2), we used the \texttt{ip\_address} function of \texttt{validators}~\cite{validators} package of Python.

\uline{Step 3.1.4 Routable IP Address}: Since assets with public IP addresses are easier to access by attackers than non-routable addresses such as localhost or private IP addresses (E-Pattern 2), we checked whether the IP addresses from Step 3.1.3 are routable. We used Python's \texttt{ipaddress}~\cite{ipaddress} package that provides functions for detecting the routable addresses.

\uline{Step 3.1.5 Scannable IP Address}: We observed that not all the public IP addresses identified from the source code are scannable (E-Pattern 3). To detect if the IP addresses from Step 3.1.4 are scannable, we did not use \texttt{ping} command since ping uses Internet Control Message Protocol (ICMP) packets that are typically blocked by servers through firewalls. In addition, ping does not provide information on the active services running on the server. Thus, we used Censys Search API~\cite{durumeric2015search}, which uses TCP and UDP packets in the network scan and maintains a database of publicly available information on the active services of a server. Finally, we filtered the scannable IP addresses and detected corresponding active service ports.

\uline{Step 3.1.6 Port Open}: Developers put the database port number in the asset identifier (E-Pattern 4). Thus, we checked whether the port is open for the scannable IP address using the open ports for the scannable IP address found in Step 3.1.5.

\begin{figure}[!t]
\centering
    \includegraphics[scale=0.58]{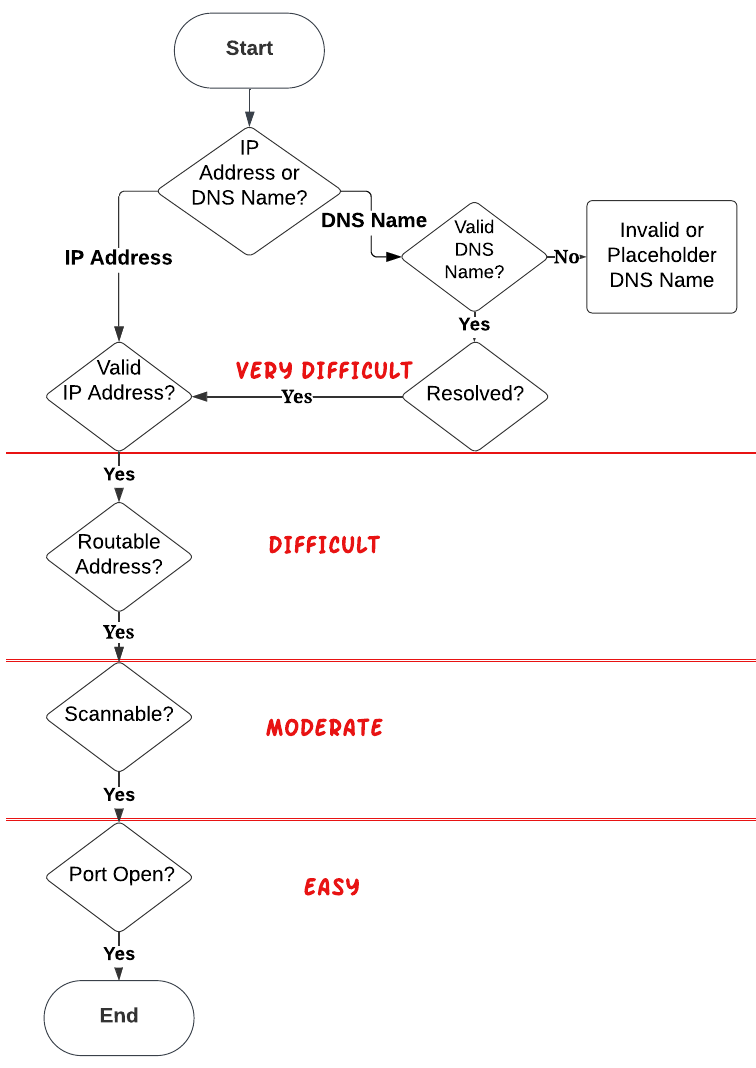}
    \caption{A flow diagram for assigning ease of attack category for an asset identified in the source code.}
    \label{fig:ease-of-attack}  
\end{figure}

\textbf{Step 3.2 Assigning Ease of Attack Category}: From Step 3.1, we observed that at each step, we find new information regarding the ease of attack of the identified asset. Thus, we need to assign an ease of attack category based on the asset information similar to the value of asset to calculate the security risk score (Step 4). To systematically assign an ease of attack category to an asset, we started with a value of 0 for ease of attack. Next, when we retrieve new information after each step, such as if the DNS name is valid (Step 3.1.1), we increment the value for ease of attack by 1. Similarly, if the valid DNS name is resolvable (Step 3.1.2), we increment the value again by 1. In our study, for ease of attack, we assigned four categories (VERY\_DIFFICULT, DIFFICULT, MODERATE, and EASY). The first and second authors of the paper independently inspected the calculated value for ease of attack and assigned a category based on the asset information. The paper's third author, who has over 15 years of experience in network security, resolved the disagreements related to the assigned categories between the first and second authors. Figure~\ref{fig:ease-of-attack} depicts the final categories assigned for ease of attack on an asset at different steps. For example, the ease of attack for an asset is MODERATE if the IP address is scannable, whereas EASY if the database port is open. Finally, we integrated the ease of attack mappings based on host information into RiskHarvester, eliminating manual effort for tool users.

\subsection{Step 4: Calculating Security Risk Score}

We identified the value of asset (Step 2) and ease of attack (Step 3) as ordinal categories. To calculate the security risk score (Equation~\ref{eqriskscore}), we need to perform ordinal scaling~\cite{agresti2012categorical}, which assigns numerical values to the categories while preserving their inherent order. Thus, to assign numerical values, we leveraged Protection Poker~\cite{williams2010protection}, a threat modeling game for security risk quantification. We conducted a Protection Poker session with the first, second, and third authors of the paper. We leveraged the nine values from 1, 2, 3, 5, 8, 13, 20, 40, and 100 used by Protection Poker for estimating the ``value of asset'' and ``ease of attack''. We assigned numerical values to the categories of value of asset and ease of attack after two Protection Poker rounds. For value of asset, the assigned values are HIGH (100), MODERATE (40), LOW (5), and UNSPECIFIED (1). For ease of attack, the assigned values are VERY\_DIFFICULT (1), DIFFICULT (8), MODERATE (40), and EASY (100). This mapping is integrated into RiskHarvester to automatically calculate the security risk score. Finally, we multiplied the value of asset and ease of attack to calculate the security risk score (Equation~\ref{eqriskscore}). For example, if the value of asset is HIGH (100) and the ease of attack is DIFFICULT (8), the security risk score is 800.

\section{Results} \label{Results}
In this section, we answer RQ1 by evaluating the performance of RiskHarvester against RiskBench and RQ2 by assessing whether developers prioritize secret removal from software artifacts ranked by descending security risk scores.

\subsection{Performance of RiskHarvester} \label{RiskHarvesterResult}

\begin{table*}[!ht]
\setlength{\tabcolsep}{3.5pt}
\centering
\caption{Precision, Recall, and F1-Score of RiskHarvester in identifying the database name, table, and column names}
\label{database-keyword-result}
\small
\begin{tabular}{|l|lll|lll|lll|}
\hline
\multicolumn{1}{|c|}{} &
  \multicolumn{3}{c|}{\textbf{Database Name}} &
  \multicolumn{3}{c|}{\textbf{Table Name}} &
  \multicolumn{3}{c|}{\textbf{Column Name}} \\ \cline{2-10} 
\multicolumn{1}{|c|}{\multirow{-2}{*}{\textbf{\begin{tabular}[c]{@{}c@{}}Database \\ Type\end{tabular}}}} &
  \multicolumn{1}{c|}{\begin{tabular}[c]{@{}c@{}}Precision \\ (TP, FP)\end{tabular}} &
  \multicolumn{1}{c|}{\begin{tabular}[c]{@{}c@{}}Recall \\ (TP, FN)\end{tabular}} &
  \multicolumn{1}{c|}{F1} &
  \multicolumn{1}{c|}{\begin{tabular}[c]{@{}c@{}}Precision \\ (TP, FP)\end{tabular}} &
  \multicolumn{1}{c|}{\begin{tabular}[c]{@{}c@{}}Recall \\ (TP, FN)\end{tabular}} &
  \multicolumn{1}{c|}{F1} &
  \multicolumn{1}{c|}{\begin{tabular}[c]{@{}c@{}}Precision \\ (TP, FP)\end{tabular}} &
  \multicolumn{1}{c|}{\begin{tabular}[c]{@{}c@{}}Recall \\ (TP, FN)\end{tabular}} &
  \multicolumn{1}{c|}{F1} \\ \hline
\textbf{MySQL} &
  \multicolumn{1}{l|}{0.98 (707, 13)} &
  \multicolumn{1}{l|}{0.96 (707, 29)} &
  0.97 &
  \multicolumn{1}{l|}{0.97 (320, 10)} &
  \multicolumn{1}{l|}{0.90 (320, 36)} &
  0.93 &
  \multicolumn{1}{l|}{0.95 (813, 46)} &
  \multicolumn{1}{l|}{0.85 (813, 133)} &
  0.89 \\ \hline
\textbf{PostgreSQL} &
  \multicolumn{1}{l|}{0.96 (598, 20)} &
  \multicolumn{1}{l|}{0.95 (598, 33)} &
  0.95 &
  \multicolumn{1}{l|}{0.96 (327, 13)} &
  \multicolumn{1}{l|}{0.94 (327, 20)} &
  0.95 &
  \multicolumn{1}{l|}{0.90 (310, 33)} &
  \multicolumn{1}{l|}{0.87 (310, 47)} &
  0.88 \\ \hline
\textbf{MongoDB} &
  \multicolumn{1}{l|}{0.90 (95, 10)} &
  \multicolumn{1}{l|}{0.82 (95, 21)} &
  0.86 &
  \multicolumn{1}{l|}{0.92 (25, 2)} &
  \multicolumn{1}{l|}{0.76 (25, 8)} &
  0.83 &
  \multicolumn{1}{l|}{0.77 (87, 25)} &
  \multicolumn{1}{l|}{0.88 (87, 11)} &
  0.82 \\ \hline
\textbf{SQL Server} &
  \multicolumn{1}{l|}{1.00 (5, 0)} &
  \multicolumn{1}{l|}{0.62 (5, 3)} &
  0.76 &
  \multicolumn{1}{l|}{1.00 (5, 0)} &
  \multicolumn{1}{l|}{0.33 (5, 10)} &
  0.51 &
  \multicolumn{1}{l|}{0.85 (12, 2)} &
  \multicolumn{1}{l|}{0.40 (12, 18)} &
  0.54 \\ \hline
\textbf{Overall} &
  \multicolumn{1}{l|}{\cellcolor[HTML]{C0C0C0}0.97 (1405, 43)} &
  \multicolumn{1}{l|}{\cellcolor[HTML]{C0C0C0}0.94 (1405, 86)} &
  \cellcolor[HTML]{C0C0C0}0.95 &
  \multicolumn{1}{l|}{\cellcolor[HTML]{C0C0C0}0.96 (677, 25)} &
  \multicolumn{1}{l|}{\cellcolor[HTML]{C0C0C0}0.90 (677, 74)} &
  \cellcolor[HTML]{C0C0C0}0.93 &
  \multicolumn{1}{l|}{\cellcolor[HTML]{C0C0C0}0.92 (1222, 106)} &
  \multicolumn{1}{l|}{\cellcolor[HTML]{C0C0C0}0.85 (1222, 209)} &
  \cellcolor[HTML]{C0C0C0}0.88 \\ \hline
\end{tabular}%
\end{table*}

\textbf{Performance of Finding Database Keywords}: Table~\ref{database-keyword-result} presents RiskHarvester's precision, recall, and F1-score in identifying the database, table, and column names for each database type. The column ``Precision (TP, FP)'' denotes the precision score with the number of true positive and false positive database keywords outputted by RiskHarvester. The column ``Recall (TP, FN)'' denotes the recall score with the number of outputted true positive and false negative database keywords. The column ``F1'' denotes the F1-score (the harmonic mean of precision and recall).


We observed that RiskHarvester demonstrated overall precision of 97\%, 96\%, and 92\% in identifying the database, table, and column names, respectively, indicating high precise detection of database keywords. The count of false positives indicates that the tool incorrectly outputted 43 database names, 24 table names, and 106 column names out of 3,304 database keywords. In addition, RiskHarvester demonstrated an overall recall of 94\% and 90\%, indicating a strong ability to identify database and table names, respectively, supported by F1-scores of 95\% and 93\%. However, the recall of identifying column names is 85\%, which is relatively lower than that of database and table names. The count of false negatives indicates that the tool failed to detect 86 database names, 74 table names, and 209 column names. We also observed that among the four database types, the recall of database, table, and column names of SQL Server is low (62\%, 33\%, and 40\%, respectively), though the precision is 100\%. In addition, the recall for table names in MongoDB is relatively low (76\%) compared to MySQL and PostgreSQL. We now discuss our observations on the false positives and false negatives.

\uline{Analysis of False Positives}: Since the SQL drivers use raw SQL queries (V-Pattern 1), we observed that the false positives on table and column names are mostly caused by the dynamically constructed queries (61\% of the false positives). We identified all the string parts flowing in the driver function sink for a SQL query using data flow analysis and reconstructed the query using the source code line and column information (Step 2.1). However, we could not reconstruct the complete query due to the presence of conditional statements and dynamically fetched values from the environment variables or config files. As a result, we identified incorrect table and column names while parsing the SQL query. Similar to dynamic raw SQL query, we observed that 24\% of the total incorrect column names are from dynamically constructed dictionary objects for column names passed in the NoSQL drivers (V-Pattern 2). Additionally, the false positives of database names are mostly triggered by the neighboring lines rule (Step 1.3), comprising 71.5\% of the 43 false positives. We observed that the prefix match of the neighboring key names met the threshold, though the key name is not the correct asset of the corresponding secret containing the database name.

\uline{Analysis of False Negatives}: We observed that the repositories of RiskBench also contain non-Python source codes such as C\# and Java. While we detected secret-asset pairs in non-Python code using regex (Step 1.1), we could not identify table and column names since data flow analysis was only applied to Python code (Step 2.1). For example, the SQL server shows a relatively lower recall (33\% and 40\% for table and column names) since SQL Server table and column names are typically passed to .NET driver functions. In addition, we reconstructed the raw SQL query from the query parts flowing into the driver sinks. However, similar to false positives, we missed table and column names due to improperly reconstructing the original query for having dynamic behavior. Additionally, we observed that 54 (4.2\%) instances of secret-asset pairs in RiskBench do not fall within three neighboring lines. Thus, we failed to detect the database name when the asset identifier containing the database name was not present in the three neighboring lines of the secret (Step 1.3).

\textbf{Performance of Sensitive Data Category Mapping}: We applied lexical and semantic matching to map the identified database keywords to the corresponding sensitive data categories (Step 2.2). We observed a precision of 85\% among the 3,673 database keywords of RiskBench. We manually inspected a random sample of 50 false positive mappings. We noticed that 27 false positives are due to the Jaro-Winkler similarity, which employs prefix bias. For example, ``credit\_limit'' and ``tax\_rate'' keywords are wrongly mapped to ``CREDIT\_CARD\_NUMBER'' and ``TAX\_ID'' data categories, respectively. In addition, we observed that 16 false positives were due to semantic matching. For example, ``transaction\_code'' is mapped to ``CREDIT\_CARD\_NUMBER'' since both terms appear in financial contexts, leading to a semantic link.   

\textbf{Performance of Detecting Placeholder Host}: We observed that the precision of identifying the placeholder host is 96\%. All the 11 false positives are DNS names outputted by the ChatGPT model (Step 3.1.1). For example, ``gg-is-awesome-246.mongodb.net'' is termed a placeholder due to the ``is-awesome'' substring in the DNS name. In addition, our tool shows a recall of 94\% for detecting the placeholder host out of 317 placeholder hosts in RiskBench. Similar to false positives, all the missing placeholder hosts are DNS names.

\subsection{Developer Survey} \label{DeveloperSurvey}

We received 52 responses (10.4\%) out of 500 developers. We now discuss our observations from the responses.


\textbf{Q1: Secret and Severity Information}: We observed that 41 developers (78\% of respondents) wanted to remove Secret A first, terming Secret B and C as placeholder/dummy, supporting our hypothesis. For example, <P23> stated \textit{``A first, then B or C. A appears to have an actual password, whereas B and C are just placeholders.''} In addition, we observed that the severity information did not help the developers. <P13> stated that \textit{``Severity info did not help, needed to look at the secret to determine that B and C are likely fake secret values.''} Additionally, 5 developers were unsure about the order of secret removal due to missing secret contexts, such as the asset information. For example, <P45> stated that \textit{``I have no idea in what order to prioritize. To effectively prioritize, I need to know the context for what these secrets grant access.''} However, 6 developers considered the asset information into account by inspecting the source code that we did not provide.


\textbf{Q2: Additional Asset Information}: We observed that 38 of 41 developers who selected Secret A in Q1 changed their priority after considering the asset identifier information. These developers changed their priority to Secrets B and C even though the secrets looked like placeholders, supporting our hypothesis. <P23> stated that \textit{``Since Secret A coming from localhost, we might not access it directly. But the other two seem on the internet and should be our top priority to address.''} However, 3 developers did not change the priority without providing any reason. Additionally, all the 5 developers who were unsure about which secrets to prioritize in Q1 have used the asset information to make decisions. <P45> stated that \textit{``Based on the added information of ip address of the system secret is used to access, I would deprioritize Secret A compared to the other two, as localhost is more likely to be hardened against outside access.''} Since 6 developers in Q1 already considered the asset information, their priority stayed the same. <P5> stated that \textit{``No, and I detailed in my previous explanation since I already took the IPs into account.''}


\textbf{Q3: Additional Security Risk Score Information}: We observed that 86\% (45 out 52) of the respondents changed their priority to Secret C, A, and B based on the descending security risk score, thus supporting our hypothesis. <P9> stated that \textit{``I would make changes to my prioritization (Secret C, A, B). Secret C has high risk score, personal data exposure, and reachable IP make it most critical to address. While Secret A has lower risk (100), high value of blockchain data means cannot be ignored, even though protected by localhost.''} Additionally, developers pointed out that they had not considered the value of asset information before. <P10> stated that ``\textit{I didn't check the Value of Asset. More security is needed for valuable assets.''} However, 4 developers did not change their priority without specifying any reason, and 3 developers wanted more context on the value of asset and ease of attack. 

\textbf{Q4: Feedback on Security Risk Score}: Developers provided feedback on the security score calculation, such as <P3> stating that \textit{``This simple calculation makes sense and is easy to understand.''} Another developer <P11> stated that \textit{``It aligns with some of the ways we do it in my sector (cloud security) at least.''} However, developers also suggested improvements to the security score calculation based on active network analysis. For example, <P40> suggested that \textit{``It would also be important to include deployment information as passive network analysis might not give the real picture.''} In addition, developers suggested accounting for whether the data is encrypted in the value of asset. <P30> stated that \textit{``If it is encrypted, it should be scored less than non-encrypted data.''} Developers also suggested including the attack vectors, such as privileges required and lateral movement, in the ease of attack calculation. <P7> stated that \textit{``attach attack vectors if possible such as privileges required. In general, attack vector score would change the prioritization.''}


\section{Discussion} \label{Discussion}
In this section, we discuss the implications of RiskHarvester based on our study findings.

\textbf{Only the asset identifier is not enough to aid developers in prioritizing the software secret removal.} Basak et al.~\cite{assetharvester} constructed AssetHarvester for detecting the corresponding asset identifier by the secret. From our developer survey, we observed that 73\% of developers changed their priority based on asset identifiers, and 10\% of developers were unsure of their decision and wanted more asset context. However, when we provided the security risk scores with the value of asset and ease of attack information, 86\% of the developers changed their priority in the descending order of security risk score. In addition, developers pointed out that the value of asset and ease of attack information helped them to make informed decisions to tackle the secret removal efforts.

\textbf{The risk-based analysis for secrets should be integrated into the secret detection tools.} We developed RiskHarvester to automatically calculate the security risk for the checked-in secrets. Our approach eliminates the need for developers to manually analyze each secret detection tool alert and calculate the security risk. To integrate our approach, the input for RiskHarvester will be the repository source code, and RiskHarvester will output the secrets ranked by descending security risk score. Thus, developers can focus their mitigation efforts on the most critical security risks.

\textbf{RiskHarvester can be extended to calculate the security risk score of secrets in other programming languages and secret types.} In our study, we calculated the security risk score of four database providers in Python. We now discuss the effort needed and challenges to extend RiskHarvester for other programming languages and secret types.

\uline{Programming Language}: We identified the secret-asset pairs using pattern matching, data flow analysis, and fast-approximation heuristics. We parsed the database names from the identified asset identifiers. Since pattern matching and fast-approximation heuristics are programming-language agnostic, we can apply the techniques in other programming languages without additional effort. Additionally, we leveraged data flow analysis to detect the secret-asset pair instance that flows into query functions and then extracted database keywords from raw SQL and ORM classes (Step 2.1). Though data flow analysis is programming language dependent, we can compute the abstract syntax tree, control flow, and data flow graph for each programming language in a repository separately using CodeQL. Next, we can identify the secret-asset pair sources and sinks from the computed graphs with minimal effort. Additionally, SQL parsing is programming-language agnostic, enabling the parsing of database keywords in other languages.

\uline{Non-database Secret Types}: From SecretBench~\cite{secretbench}, we inspected five random samples of secrets of seven secret types, such as API keys, private keys, and authentication tokens. We now discuss extending RiskHarvester to identify non-database secret-asset pairs and corresponding asset keywords.

\textbf{1. Secret-Asset Pairs}: The 2024 GitGuardian report~\cite{gitguardian-secret-sprawl} reveals that cloud secrets such as API keys and tokens are the second most exposed in GitHub. Since cloud providers have specific formats, we can identify the secret-asset pair using the regex (Step 1.1). Additionally, we can identify the functions of frameworks such as .NET and Spring, where the secret-asset pairs are passed similarly to database drivers and employ data flow analysis (Step 1.2). The list of functions will not be huge since most non-database secret-asset pairs are passed in HTTP clients such as \texttt{get} and \texttt{post} functions. 

\textbf{2. Non-Database Keywords}: We inferred the asset's value from raw SQL and ORM parsing for database secrets. However, for non-database secrets, we can infer the data category from the request body and response of HTTP requests that use the secret-asset pair. Thus, using data flow analysis (Step 2), we can identify the request body (sinks) and parse the request body parameters (sources) to infer the data categories. However, the responses of HTTP requests (typically in JSON or XML format) are serialized into classes. Thus, we can employ data flow analysis to detect the response class and identify the data categories from the class attributes. In our study, we used the \texttt{py-models-parser}~\cite{py-models-parser} package to parse ORM models (Step 2.1), which also supports parsing any data class.


\textbf{The security risk score can be improved by employing active network analysis in RiskHarvester.} Our study focused on passive network information for estimating the ease of attack. However, developers wanted the deployment-related security information, such as the level of security controls present on the asset and network vulnerabilities (Section~\ref{DeveloperSurvey}). We will extend RiskHarvester by leveraging the information provided by the active network scanning and vulnerability assessment tools such as Nessus~\cite{nessus} and Nmap~\cite{nmap}. Additionally, we will improve the value of asset by retrieving the metadata information of a database, such as whether the data is encrypted or hashed, using database auditing tools such as Datadog~\cite{datadog}. We will deploy RiskHarvester in the organization where these scanning and auditing tools are deployed.

\section{Threats to Validity} \label{ThreatToValidity}
In this section, we discuss the limitations of our study. 

\textbf{Manual Analysis}: We identified database keywords and the data categories for each secret-asset pair of RiskBench by manually inspecting the source code (Section~\ref{RiskBench}). However, manual analysis is prone to bias due to differing interpretations and oversights. Two authors cross-checked the identified database keywords and data categories to mitigate the bias. 

\textbf{Benchmark Dataset}: Our benchmark dataset selection is susceptible to bias. Basak et al.~\cite{assetharvester} identified the secret-asset pairs of AssetBench using two open-source tools, TruffleHog and Gitleaks, from GitHub repositories. However, these two tools may miss secrets from the repositories. Additionally, AssetBench does not contain repositories from other VCSs, such as BitBucket~\cite{bitbucket}. We could not mitigate the potential bias since AssetBench is the only publicly available dataset.

\textbf{Developer Survey}: Our survey findings are susceptible to external validity, as the participant pool might not accurately represent the broader developer population. To mitigate the limitation, we randomly sampled developers with prior experience in software secrets. The survey results may be influenced by how we presented the problem to the developers. To ensure clarity in the survey questions, we conducted a pilot survey with security researchers and refined the questions based on their feedback. Additionally, we provided open-ended questions to mitigate the bias from predefined options~\cite{tourangeau2000psychology}.


\section{Related Work} \label{RelatedWork}
Prior researchers~\cite{meli2019bad,rahman2019share,rahman2019seven, igibeksecret, rahman2021different, krause2023pushed} studied the root causes of secret exposure and found that keeping hard-coded secrets in software artifacts as the most prevalent insecure practices among developers, leading to secret leaks. Meli et al.~\cite{meli2019bad} found over 100K hard-coded secrets by studying a 13\% snapshot of GitHub repositories in 2019. Within Infrastructure as Code (IaC) scripts, Rahman et al.~\cite{rahman2019seven}  found 7 ``Security Smells'' by studying 5,232 IaC scripts from 293 repositories. They found that hard-coded secrets are the most prevalent among the security smells, with 1,326 occurrences. Rayhanur et al.~\cite{rahman2019share} studied 5,822 Python Gists in GitHub and found 689 hard-coded secrets, thus indicating that hard-coded secrets have been leaked in various forms of software artifacts.

To prevent secret leaks in software artifacts, researchers~\cite{krause2023pushed,basaksecretpractice, basakchallenges} recommended developers follow secure practices for secret management. Basak et al.~\cite{basaksecretpractice} conducted a grey literature review of Internet artifacts in 2022 and found 24 developer and organization practices for secure secret management. They suggested using VCS scan tools to prevent accidental secret commits. They also studied the developer's challenges for checked-in secrets by analyzing 779 questions from Stack Exchange (SE) and the solutions suggested to mitigate the challenges by SE users~\cite{basakchallenges}. The SE users also suggested using VCS scan tools to prevent accidental secret commits. However, Basak et al.~\cite{basak2023compare} found VCS scan tools outputting 25-99\% false positives and missing 14-99\% of secrets of a repository by comparing 5 open-source and 4 proprietary tools against SecretBench~\cite{secretbench}. Though Machine Learning (ML) algorithms~\cite{fengsecret, secrethunter, konygin2023using} have been used to reduce false positives, Basak et al.~\cite{basak2023compare} found 2 tools (Commercial X and SpectralOps~\cite{spectralops}) employing ML among the 9 tools showed lower precision of 25\% and 1\%, respectively. Additionally, Rayhanur et al.~\cite{rahman2022secret} found developers ignoring secret alerts due to high false positives and time pressure through a developer survey in a company (anonymized). Basak et al.~\cite{assetharvester} developed AssetHarvester, a static analysis tool to detect asset identifiers protected by secrets. Since each secret-asset pair poses a different security risk, we built upon their work and studied to provide security risk scores for secret-asset pairs to aid developers in prioritizing secret removal. 

\section{Conclusion} \label{Conclusion}
We present RiskHarvester, a risk-based tool to compute a security risk score based on the value of the asset and ease of attack on a database. We calculated the value of asset by identifying the sensitive data categories present in a database from the database keywords. We utilized data flow analysis, SQL, and Object Relational Mapper (ORM) parsing to identify the database keywords. To calculate the ease of attack, we utilized passive network analysis to retrieve the database host information. To evaluate RiskHarvester, we curated RiskBench, a benchmark of 1,791 database secret-asset pairs with sensitive data categories and host information manually retrieved from 188 GitHub repositories. RiskHarvester demonstrates precision of (95\%) and recall (90\%) in detecting database keywords for the value of asset and precision of (96\%) and recall (94\%) in detecting valid hosts for ease of attack. Finally, we conducted an online survey to understand whether developers prioritize secret removal based on security risk score. We found that 86\% of the developers prioritized the secrets for removal with descending security risk scores.

\section{Ethics Considerations} \label{Ethics}
In our study, we followed the USENIX Ethics Guidelines~\cite{ethics-guideline}. Since RiskBench contains sensitive information, the dataset will be selectively shared with researchers and tool developers under a data protection agreement to ensure ethical use. Additionally, we obtained IRB approval from our institution (Uni IRB blinded) before conducting the survey. No personally identifiable information was collected from participants apart from the email addresses of the participants who wished to participate in the lottery. Additionally, a consent form was included in the email stating that participants should not attempt to use the secret-asset pairs to verify their validity. We have also stated in the consent form that participation was voluntary and participants could withdraw at any time. 

\section{Open Science} \label{OpenSciencePolicy}
Our curated dataset, RiskBench, is stored in Google BigQuery (Dataset ID: \textit{\seqsplit{dev-range-411201.riskbench}}) as a relational structured data. Researchers and tool developers can utilize and extend the dataset for future research using SQL queries in Google BigQuery. Additionally, we have made the implementation of RiskHarvester publicly available~\cite{riskartifacts}.





\bibliographystyle{plain}
\bibliography{bibliography}

\appendix
\begin{table*} [!htb]
\footnotesize
\centering
\caption{Regexes categorized into three groups based on connection string format similarity for identifying secret-asset pairs}
\label{regex-database-appendix}
    \includegraphics[width=\textwidth]{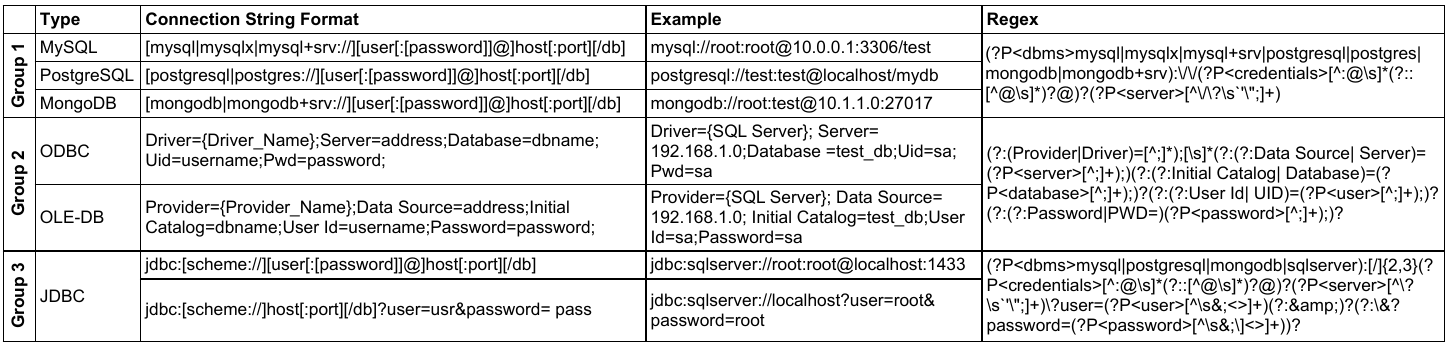}
\end{table*}

\begin{table*}[]
\centering
\caption{System and User role prompt for detecting placeholder/dummy DNS name.}
\label{dns-prompt}
\small
\begin{tabular}{|ll|l|}
\hline
\multicolumn{2}{|c|}{\textbf{Type}} &
  \multicolumn{1}{c|}{\textbf{Chain-of-Thought Prompting}} \\ \hline
\multicolumn{2}{|l|}{System} &
  \begin{tabular}[c]{@{}l@{}}In source code, developers sometimes use placeholder/dummy DNS names instead of actual DNS names. \\ For example,  in the code snippet below, "www.example.com" is a placeholder/dummy DNS name.\\ \\ -- Start of Code --\\ mysqlconfig = \{\\      "host": "www.example.com",\\      "user": "hamilton",\\      "password": "poiu0987",\\      "db": "test"\\ \}\\ -- End of Code -- \\ \\ On the other hand, in the code snippet below, "kraken.shore.mbari.org" is an actual DNS name.\\ \\ -- Start of Code --\\ export DATABASE\_URL=postgis://everyone:guest@kraken.shore.mbari.org:5433/stoqs\\ -- End of Code -- \\ \\ Given a code snippet containing a DNS name, your task is to determine whether the DNS name is a placeholder/dummy name. \\ Output "YES" if the address is dummy else "NO".\end{tabular} \\ \hline
\multicolumn{2}{|l|}{User} &
  \begin{tabular}[c]{@{}l@{}}Is the DNS name "\{dns\}" in the below code a placeholder/dummy DNS? \\ Take the context of the given source code into consideration.\\ \\ \{source\_code\}\end{tabular} \\ \hline
\end{tabular}%
\end{table*}

\end{document}